\documentclass[runningheads]{llncs}
\usepackage{graphicx}
\usepackage{xspace}
\usepackage{amssymb}
\usepackage{booktabs}
\usepackage{colortbl}
\usepackage[x11names,dvipsnames,table]{xcolor}
\usepackage{cite}
\usepackage{url}
\usepackage{breakurl}

\begin{document}
\newcommand{\model}{TUEF\xspace}
\title{Towards a Robust Expert Finding in Community Question Answering Platforms}
%
%
\author{Maddalena Amendola\inst{1, 2}\orcidID{0000-0001-6556-4032} \and
Andrea Passarella\inst{2}\orcidID{0000-0002-1694-612X} \and
Raffaele Perego\inst{3}\orcidID{0000-0001-7189-4724}}
\authorrunning{M. Amendola et al.}
%
\institute{University of Pisa, Italy \and
IIT-CNR, Italy \and
ISTI-CNR, Italy 
}
\maketitle              
\begin{abstract}
This paper introduces \model, a topic-oriented user-interaction model for fair Expert Finding in Community Question Answering (CQA) platforms. The Expert Finding task in CQA platforms involves identifying proficient users capable of providing accurate answers to questions from the community. To this aim, \model improves the robustness and credibility of the CQA platform through a more precise Expert Finding component. The key idea of \model is to exploit diverse types of information, specifically, content and social information, to identify more precisely experts thus improving the robustness of the task. We assess \model through reproducible experiments conducted on a large-scale dataset from StackOverflow. The results consistently demonstrate that \model outperforms state-of-the-art competitors while promoting transparent expert identification.

\end{abstract}

\section{Introduction}
The Expert Finding (EF) task in Community Question Answering (CQA) platforms aims to identify and recognize community users with a high level of expertise. Newly posted questions are forwarded to them, reducing the waiting time for answers. EF is crucial for enhancing the platform's overall quality, as it determines, by and large, the credibility and trustworthiness of the platform. Precise identification of experts is a key feature in guaranteeing high-quality answers and the credibility of the CQA platform.
Given a question posted by a user of the online community, the EF task can be cast directly to the task of \textit{computing a short, ranked list of community members, i.e., experts, that are likely to provide an accurate answer to the question.}

\begin{figure}
\vspace{-12pt}
    \centering    
    \includegraphics[width=1\textwidth]{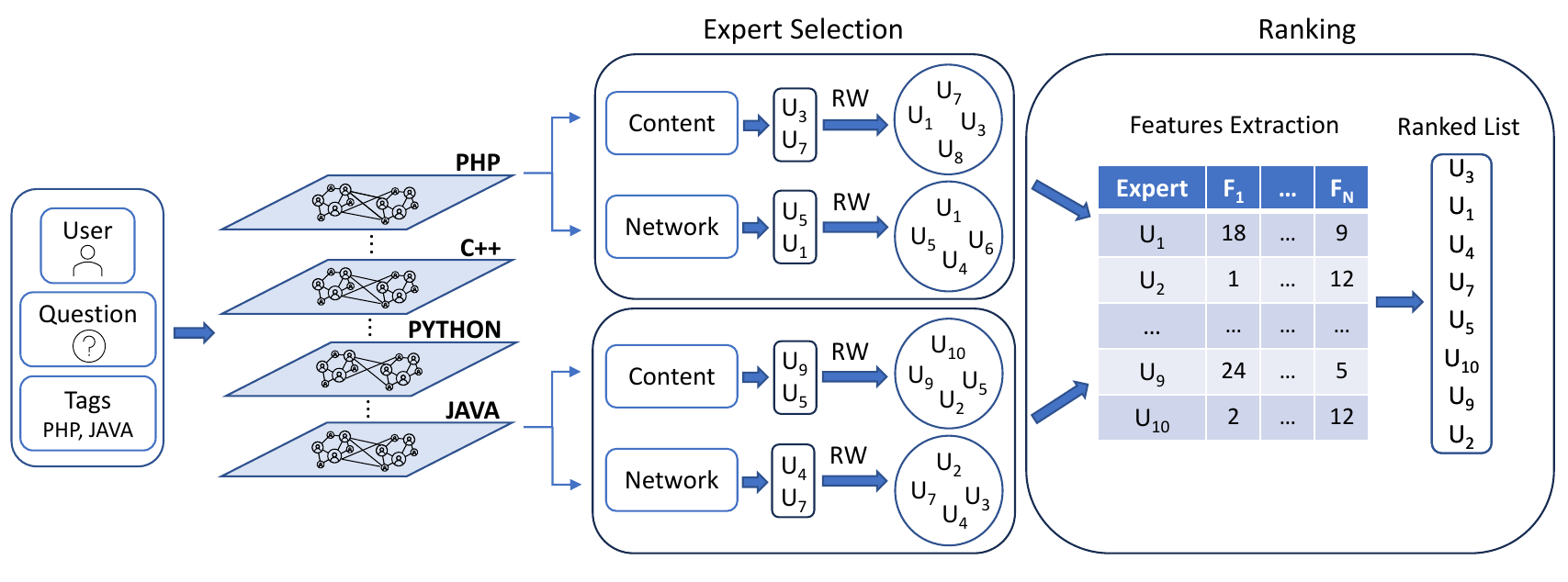}
    \caption{\model approach representation highlighting the distinct components.}
    \label{fig:tuef}
    \vspace{-12pt}
\end{figure}

Numerous studies have suggested solutions for the EF task that rely solely on textual information. Nevertheless, CQA platforms provide information through which an implicit social network among users can be defined and exploited to identify experts. Exploiting multiple sources of information (such as content and social relationships) is also a way to make the EF task more robust, as algorithms can be built on complementary and largely orthogonal sources of information. 
\indent Based on these considerations, this paper proposes \model, a \textit{Topic-oriented, User-interaction model for EF}. TUEF jointly leverages \textit{content} and \textit{social} information available in the CQA by defining a topic-based Multi-Layer graph (MLG) that represents CQA users' relationships based on their similarities in providing answers. 
TUEF stands out for its integrated approach, blending social network analysis with information retrieval techniques. One of its notable features is its transparency. Unlike deep neural network approaches that operate as black boxes, \model's approach ensures a transparent expert selection process: the method's graph exploration begins from key nodes, ensuring a straightforward selection process for candidate experts. Further enhancing its performance, TUEF employs LambdaMart \cite{burges2010ranknet}, a decision tree-based learning-to-rank method. This not only refines the ranking of experts but also showcases the contribution of the various features to the decision process, allowing for insights into the importance of various factors in the ranking outcome.
Figure \ref{fig:tuef} outlines the main building blocks of the proposed solution. \model first generates an MLG, where each layer corresponds to one of the main topics discussed in the community.
In each layer, the nodes represent users actively participating in topic discussions, while edges model similarities and relationships among users under specific topics. At inference time, given a question $q$, the MLG $G$, and a ranking model $r$, \model first determines the main topics to which the question $q$ belongs and the corresponding graph layers. Next, for each layer, it selects the candidate experts from two perspectives: i) \textit{Network}, by identifying central users that may have considerable influence within the community;
ii) \textit{Content}, by identifying users who previously answered questions similar to $q$. In both cases, the graph is used to collect candidate experts through appropriate exploration policies. Then, \model extracts features based on text, tags, and graph relationships for the selected experts. Finally, it uses a learned, precision-oriented model $r$ to score the candidates and rank them by expected relevance.

The different components of TUEF sketched in Figure~\ref{fig:tuef} are presented in detail in Section~\ref{sec:TUEF-details}. We evaluate the performance of TUEF in terms of ranking quality metrics by comparing it with state-of-the-art proposals, as well as with reference benchmarks which only consider part of the information included in the \model model (Sections~\ref{sec:experimental setup} and \ref{sec:results}). The reproducible experiments conducted are based on a large-scale dataset available from StackOverflow\footnote{\url{https://stackoverflow.com}}, the largest community within the Stack Exchange network. Before detailing the methodology adopted for \model, we discuss related work in Section~\ref{sec:related}, while in Section~\ref{sec:conclusions} we draw the main conclusion of our study.

The experimental results show that \model consistently outperforms the competitors and effectively exploits the heterogeneous sources of information modeling users and questions. Specifically, \model outperforms the best-performing competitor by 65.06\% for Mean Reciprocal Rank and 112\% for P@1.

\section{Related Work}
\label{sec:related}

Research proposals in the field of the EF task for CQA platforms can be grouped into three broad groups: text-based, feature-based, and network-based methods.
\textit{Text-based} methods address the EF task by relying on the similarities between the current and previously answered questions. 
In \cite{DBLP:conf/wsdm/ZhangCZCXLC20}, the authors propose a temporal context-aware representation learning model, which models temporal dynamics by multi-shift and multi-resolution settings. The model learns the expert representation in the context of a question's semantic and temporal information using a pre-trained deep bidirectional transformer \cite{DBLP:conf/naacl/DevlinCLT19}. In \cite{fu2020recurrent}, the authors define a new approach based on the Recurrent Memory Reasoning Network (RMRN), composed of different reasoning memory cells that implement attention mechanisms to focus on different aspects of the question. Recently, Liu et al. \cite{liu2022efficient} proposed a Non-sampling Expert Finding model that could learn accurate representations
of questions and experts from whole training data. Other similar approaches include \cite{DBLP:journals/tkdd/DehghanA19,DBLP:conf/www/Liang19,DBLP:journals/ipm/DehghanBA19}

\textit{Feature-based} methods methods are based on a set of hand-crafted features modeling the expertise of the community members. 
In \cite{tondulkar2018get}, the authors include features that favor experts who provide few but high-quality answers to complex questions, considering a broad set of features that capture the user availability and knowledge and applying, as in \model,  Learning to Rank methods to the expert ranking task. 
In \cite{fu2020user}, the authors consider the intimacy between the asker and answerer, proposing a User Intimacy Model. Differently, Peng et al. \cite{peng2022expertplm} integrate the vote score embedding with the corresponding question embedding to model the expert ability and propose a reputation-augmented Masked Language Model (MLM) pre-training strategy to capture the expert reputation information.
Other similar approaches include \cite{faisal2019expert,roy2018finding, mumtaz2019expert2vec, fu2019tracking, kundu2019finding}.

Finally, \textit{network-based} methods integrate network, interactions, and relationship information. Kundu et al. \cite{kundu2019formulation} define a framework that includes a text-based component, which estimates an expert knowledge of a topic, and a Competition Based Expertise Network \cite{aslay2013competition} that exploits link analysis techniques \cite{liu2005co, li2002improvement}. 
Differently, Sun et al. \cite{DBLP:conf/icwsm/SunMR018} follow the idea that a user's expertise is language-agnostic. The authors build a competition graph containing users and question nodes, with the edges representing relationships among them, capable of encoding the hierarchical structure of questions. 
Other approaches in this class are presented in \cite{kundu2020preference, le2018retrieving, DBLP:conf/aaai/SunBBLS019, li2019personalized}.

\paragraph{\textbf{Position of our work.}}\model can be considered a network-based, holistic model that jointly integrates text, community, and social information. It models user relationships under community topics with an MLG. It applies Information Retrieval and Social Network analysis techniques to identify the candidate experts to forward the new question. Finally, it uses a Learning-to-Rank model to precisely order these candidates.

\section{\model design}
\label{sec:TUEF-details}

In typical CQA, users tag their questions to characterize them. In addition, it is possible to exploit users' behavior to derive implicit "social" relationships based on the similarity of users' behaviors. To effectively incorporate the valuable information derived from tags, content, and user relationships, \model employs an MLG (Section \ref{sec:mlg}) representing the macro topics discussed in the CQA platform.
Each graph layer models the users' relationships at the level of the specific macro topic by considering the users' similarities as captured by tags, questions, and answering behaviors. \model adopts an exploratory algorithm that comprehensively takes full advantage of the MLG structure to select a set of candidate experts, i.e., the users who have earned a reputation by consistently providing accepted answers to questions, for each macro topic the current question belongs to (Section \ref{sec:selection}). The exploratory algorithm jointly leverages \textit{social} and \textit{content} information, considering experts' centrality in the network and the users' expertise in the topics of the question, increasing the likelihood of selecting high-quality experts. The selection process is inherently transparent, guided by the graph structure that delineates user relationships.
Finally, \model extracts features representing the identified experts and applies Learning-to-Rank (LtR) techniques (Section \ref{sec:ranking}) to sort them according to their expertise and likelihood of answering the question. \model uses a decision tree-based LtR approach providing a high level of clarity on the significance of different factors influencing the ranking result \cite{10.1145/3477495.3531840}.

\subsection{User Interaction Model}
\label{sec:mlg}
Hereinafter, let $U$ be the set of active users in the CQA platform considered, and  $Q$ the set of historical questions posted and answered by users in $U$.
\vspace{-6pt}
\subsubsection{Topic Identification}
\label{sec:clustering}
\model clusters the tags associated with the historical questions posted on the CQA platform by adopting the solution discussed in \cite{clustering}, which applies the \textit{k}-means algorithm to a tag co-occurrence matrix $M$. 
Tags' co-occurrence patterns provide helpful information for clustering since tags often associated with the same questions are likely semantically related \cite{4597030}. 
Moreover, the adopted methodology is based on the observation that CQA users often use very common and broad tags paired with more specific ones to ensure greater precision in topic identification. We can enhance the questions' categorization by considering the co-occurrences of discriminating tags with the broader ones.

Let $T=\{t\;|\;t\in tags(q),\;q\in Q\}$ be the set of all tags associated with questions in $Q$, where $tags(q)$ is the tags list of the question $q$, and $F$ the set of the $\lambda$ most frequent tags occurring in questions of $Q$ representing the clustering features. 
The co-occurrences matrix $M^{|T| \times |F|}$ is constructed as follows:
\begin{equation}
    m_{i, j} = |\{q \in Q \;|\; \{t_i, f_j\} \subseteq tags(q), \; t_i \in T, \; f_j \in F\}|
\end{equation}
Element $m_{i, j}$ indicates the number of questions in $Q$ where the $i_{th}$ tag and the $j_{th}$ feature co-occur. 
We obtained a normalized matrix $\hat{M}$ by normalizing $M$'s rows to express the fraction of tag co-occurrences relative to each feature.

Given a value \textit{k}, the \textit{k}-means hyperparameter, the clustering algorithm returns \textit{k} disjoint clusters of tags that represent the main domain areas of the community considered.
As detailed in Section~\ref{sec:experimental setup}, we use the silhouette maximization criteria \cite{rousseeuw1987silhouettes} to identify the proper value of \textit{k}.
\vspace{-6pt}
\subsubsection{Multi-Layer Graph}
\label{subsec:mlg}
\model models users' relationships within each layer by representing users $U$ as nodes and establishing a connection between two nodes if the corresponding users have a similar pattern of providing accepted answers to questions related to the specific layer.
Formally, \model uses a MLG $G = [L_1, ..., L_k]$ where $L_i$ represents the layer of $G$ associated with the $i-th$ tag cluster.
Each layer $L_i = (V_i, E_i)$ is an independent graph (i.e., there are no edges between nodes of different layers), where $V_i$ is the set of nodes representing users in layer $L_i$ and $E_i$ is the set of edges representing users' relationships. 

To focus on users who can consistently provide accurate answers for questions related to the specific layer, we implement a filtering mechanism that includes in $V_i$ only users that provided at least $\epsilon$ accepted answers, i.e., $V_i \subseteq U$. 

To model the users' knowledge, we build for each user $u \in V_i$ a topic vector $b^i_u$ where the $j_{th}$ position of $b^i_u$ stores the number of accepted answers provided by $u$ to questions associated with the $j_{th}$ tag of $L_i$, normalized based on the total number of accepted answers provided by the user in all the layers\footnote{When modeling the knowledge, we discard the $\lambda$ high-frequent tags, used as features during the clustering phase, to focus on more discriminating ones.}.

After representing each user $u$ of $V_i$ with the topic vector $b^i_u$, we compute the cosine similarity between all pairs of vectors within the same layer. If the similarity between two users $u_a$ and $u_b$ exceeds a predetermined threshold value $\delta$, we insert in $E_i$ an edge $(u_a, u_b)$ weighted by the cosine similarity value.

It is important to note that each question in \model is associated with a list of tags, where each tag is assigned to only one layer within the MLG $G$. Consequently, a question may belong to multiple layers within $G$. Furthermore, the users who answer these questions are represented in all the associated layers, implying that a user's expertise and social relationships span across all the layers to which the questions they answer belong. Considering the multidimensionality of user expertise across various layers, \model captures a complete representation of users' knowledge and interactions within the CQA platform.

\subsection{Expert Selection}
\label{sec:selection}
Given a new question $q$, the Expert Selection component is performed multiple times, one for each layer $q$ belongs to. The selection process is divided into three phases: the \emph{sorting} phase, which sorts the nodes in each layer; the \emph{collection} phase, which selects an initial set of experts from the sorted lists; the \emph{exploratory} phase, which expands the initial set of experts by exploring the graph.  
\vspace{-6pt}
\subsubsection{Expert Identification}
\label{sec:exp_rep}
Experts in CQA platforms are identified effectively by considering the number of \textit{accepted answers} they provided in the past. 
Another signal of user trust is given by the acceptance ratio $r_u$, i.e.,  the ratio between the number of accepted answers and the total number of answers a given user provides. 
As in \cite{dargahi2017skill}, the set $C \subseteq U$ of \textit{candidate experts} is selected by considering all the users having a number of accepted answers greater or equal to a specified threshold $\beta$. Finally, the set of \textit{experts} $E \subseteq C$ includes the candidates whose acceptance ratio $r_u$ is greater than the overall average $\overline{r}$.
Note that each layer includes \textit{experts} and \textit{non-experts} nodes. The former is the target of the Expert Selection process; the latter serves as node transit to fully explore the MLG.
\vspace{-6pt}
\subsubsection{Sorting}
To exploit content and social information, \model considers two complementary perspectives: the users' centrality within the network (\textit{Network-based} perspective) and the relevance of the user to the newly posted query $q$ based on her previously answered questions (\textit{Content-based} perspective).

The Network-based approach uses the Betweenness centrality \cite{freeman1977set}, which assesses the centrality of nodes within a graph by considering the shortest paths and representing the influence a node has over the flow of information in a graph. Central nodes will have a higher value of Betweenness centrality.
The expert nodes $v_j^i$ in the layer $L_i$ are sorted considering their centrality score $s_j^i$.

In contrast, the Content-based approach sorts the layer's expert nodes based on the similarity to the query $q$ of the questions answered in the past. To compute content-based similarity, \model rely on Information Retrieval techniques and pre-built indexes.  Specifically, it instantiate two indexes: the \textit{TextIndex} and the \textit{TagIndex}, indexing the text and tags associated with historical questions, respectively.
Given the new question with associated tags, the Content-based method uses a retrieval model (e.g., BM25) on both indexes, each returning a sorted list of questions along with the information about the experts who provided the accepted answer. For the \textit{TagIndex}, the query is composed by concatenating the question's tags, while for the \textit{TextIndex}, the query is the concatenation of the question's title and body. 
Finally, the two lists are merged by alternatively taking one item from each list while preserving the elements' original order.
The final result is a query-dependent ordering of the nodes of each layer $L_i$.  
\vspace{-6pt}
\subsubsection{Candidate collection}
Candidate collection involves selecting a subset of experts $D_i \subseteq V_i$ for each method in each layer $L_i$ the question belongs to. 
On the one hand, the objective is to consider a set of candidates large enough to achieve a high probability of obtaining the correct answer from them, i.e., high recall. Conversely, we aim to consider a set of candidates potentially including only users who are experts on the specific question, i.e., high precision. To tradeoff between precision and recall, we estimate the probability $p$ of not receiving an answer from any user in $D_i$. \model starts collecting the experts from the sorted list computed in the sorting step.
The probability $p$, initialized to 1, is incrementally updated as experts are included in $D_i$. Specifically, it is reduced as a function of $\mu_u$ corresponding to any new expert $u$ added to $D_i$, which is the ratio between the number of accepted answers over the total number of answers provided by $u$. 
Moreover, to better model the topic-based expertise, $\mu_u$ is smoothed by considering the user's activity on the specific layer $L_i$. This smoothing is computed by multiplying $\mu_u$ by the ratio between the number of expert answers in the layer and the maximum number of answers provided in $L_i$ among all users. 

The value of  $\mu_u$ estimates the probability that an expert will answer the question by modeling her capability and overall activity in the layer. The probability $p$ is thus updated as $p = p \cdot (1 - \mu_u)$, until it becomes less than or equal to a threshold $\alpha$. At this point, we start the Exploratory phase.

\vspace{-6pt}
\subsubsection{Exploratory phase}
During the Exploratory phase, \model explores the graph structure to gather additional experts by leveraging user relationships.
The experts $D_i$ from each layer $L_i$ serve as the starting nodes for the exploration.
For each node $v_i \in D_i$, \model conducts a fixed number of Random Walks and, at each step, randomly selects the next node to visit based on the probability distribution $d$ computed considering the neighbors of the current node $v_i$ and its links weights, representing the similarities with its neighbors. 
Specifically, given an expert $v_i$, we denote with $\mathcal N_i=\{v_{i1}, \ldots, v_{iN_i}\}$ the neighbours of $v_i$ (where $N_i=\left|\mathcal N_i\right|$ is the total number of $v_i$'s neighbours). The probability of visiting $v_{ij}$ is given by the ratio between its weight $w_{ij}$ and the sum of the neighbours' weights  in $N_i$. 
This process is computed independently for each layer to which the question belongs. Additionally, within each layer, Random Walks are performed for both the Network-based and Content-based methods.

\subsection{Ranking}
\label{sec:ranking}

To learn an effective ranking function from the training data we resort to \textit{Learning to Rank}  (LtR)~\cite{10.1561/1500000016}.
{LtR} algorithms exploit a {\em ground-truth} to learn a scoring function $\sigma$ mimicking the ideal ranking function hidden in the training examples.
\model extracts a subset of the historical questions used to model the users' relationships in Section \ref{sec:mlg} as the training set for the LtR algorithm.
Specifically, each query $q$ in the training set is associated with a set of candidate experts $U = \{u_{0}, u_{1}, \ldots \}$. Each query-candidate pair $(q,u_i)$ is in turn associated with a \emph{relevance judgment} $l_{i}$ establishing if $u_i$ is an expert for query $q$ or not. 
Query-candidate pairs $(q,u_i)$ are represented by a vector of features $x$,  able to describe the query, the expert, and their relationship.
The LtR algorithm learns a function $\sigma(x)$ predicting a relevance score for the input feature vector $x$. Such function $\sigma(x)$ is finally used at inference time to compute the candidate experts' scores and rank them accordingly. 

The features modeling the query and a candidate expert can be categorized into two groups: \textit{static} and \textit{query-dependent} features. The \textit{static} features model the relative importance of each expert selected during the Expert Selection process, disregarding any relevance to the specific query. They include:
(i) the \textbf{Reputation} of the expert; (ii) the total number of \textbf{Answers} and \textbf{AcceptedAnswers} provided by the expert; (iii) the \textbf{Ratio} between the Answers and AcceptedAnswers; (iv) \textbf{AvgActivity} and \textbf{StdActivity} features, which are the average and standard deviation derived from the time differences between consecutive answers provided by the expert, respectively.
In contrast, the \textit{query-dependent} features are derived from the Expert Selection process and concern the content and topics of the specific query. The \textit{query-dependent} features modelling the query $q$ and the expert $u$ are the following: (i) \textbf{LayerCount}: the number of distinct graph layers in which the expert is selected during the Expert Selection process; (ii) \textbf{QueryKnowledge}: the ratio between the number of Answers and AcceptedAnswers provided by $u$ in the layers relevant for $q$; (iii) \textbf{VisitCountContent} and \textbf{VisitCountNetwork}: the total number of times an expert is encountered in  the Collection and Exploratory phases using the Content-based and Network-based techniques, respectively; (iv) \textbf{StepsContent} and \textbf{StepsNetwork}:  the number of steps necessary to first discover $u$, either during the Collection or Exploratory phase; (v) \textbf{BetweennesPos} and \textbf{BetweennessScore}: the expert's rank in the list of users ordered by Betweenness score and the Betweenness score itself; (vi) \textbf{ScoreIndexTag} and \textbf{ScoreIndexText}: the sum of BM25 scores of historical questions answered by the experts in the IndexTag and IndexText, respectively; (vii) \textbf{FrequencyIndexTag} and \textbf{FrequencyIndexText}: the number of distinct questions answered by the expert returned by the respective indexes; (viii) \textbf{Eigenvector}, \textbf{PageRank} and \textbf{Closeness}: the scores of the Eigenvector, PageRank, and Closeness centrality measure, respectively; (ix) \textbf{Degree}: the node Degree of the expert node; and (x) \textbf{AvgWeights}: the average of the links' weights of the expert node.
Note that features like Eigenvector, PageRank, Closeness, Degree, AvgWeights, are query-dependent and not static features because the \model graph layers on which they are computed depend on the tags associated with the query. 

For each candidate expert selected in more than a layer, we consider the maximum of values for the Network-based features and the sum of values for the Content-based, \textit{LayerCount} and \textit{QueryKnowledge} features. For \textit{StepsContent}, \textit{StepsNetwork}, 
\textit{BetweennesPos} we consider the minimum value.

\vspace{-12pt}
\begin{table}[]
\caption{Statistics of the StackOverflow dataset used.}
\centering
\rowcolors{2}{gray!15}{white}
\begin{tabular}{rcccc}
 & \multicolumn{1}{r}{\textbf{Questions}} & \multicolumn{1}{r}{\textbf{Answers}} & \multicolumn{1}{r}{\textbf{Users}} & \multicolumn{1}{r}{\textbf{Tags}} \\ 
 \toprule
{Raw Dataset} & 877,180 & 1,155,189 & 618,659 & 36,759 \\ 
{Train Dataset} & 268,185 & 373,472 & 211,057 & 23,245 \\ 
{Test Dataset} & 17,002 & 21,879 & 18,181 & 6,380 \\ 
\bottomrule
\end{tabular}
\label{tab:statistics}
\vspace{-12pt}
\end{table}

\section{Experimental Setup}
\label{sec:experimental setup}

We conducted experiments using a large-scale dataset from StackOverflow, the largest community within the Stack Exchange network. The StackOverflow data dump is publicly available\footnote{\url{https://archive.org/details/stackexchange}} and contains more than 22 million questions posted by community members from 2008-07-31 up today. 

\textbf{Preprocessing} We selected the data from 2020-07-01 to 2020-12-31 to focus on a six-months time range. As in \cite{mumtaz2019expert2vec}, we followed well-established cleaning practices to ensure data quality. First, we removed all questions and answers without a specified Id and OwnerUserId. We kept only questions with an AcceptedAnswerId and answers with a specified ParentId, corresponding to the Id of the question it relates to. Finally, we removed questions where the asker and the best answerer were the same user. We divide the questions using 80\%  for training/validation and the remaining 20\% for the test set. Importantly, we preserve the temporal order of the questions. Statistics about the resulting dataset are reported in Table \ref{tab:statistics}.

\textbf{User Interaction Model} To identify the macro topics discussed in the StackOverflow community, we apply the clustering technique detailed in Section \ref{sec:clustering} to questions' tags of the training set. Considering $\lambda=10$ most frequent tags, we obtain $k=10$ clusters as the value of $k$ maximising the Silhouette score.
Each cluster represents a layer of the MLG where we define the interaction between users who gave at least $\epsilon=3$ accepted answers. We represent the users as topic vectors, as reported in Section \ref{sec:mlg}. Finally, when modeling relationships, we keep the edges with a similarity equal to or greater than $\delta=0.5$. 

\textbf{Expert Selection} We model the experts by selecting all users with at least $\beta=20$ accepted answers and following the procedure explained in Section \ref{sec:selection}, resulting in 1,230 experts.
During the expert selection process, the two indexes are queried for the 1,000 past questions most similar to the query. After sorting the nodes, the collection phase selects the initial set of experts $D$, reaching a probability $p=0.001$ representing the probability of not getting an answer. Finally, we perform $5$ Random Walks of $10$ steps for each selected expert. 

\textbf{Ranking} 
We use the implementation of LambdaMART available in LightGBM \cite{NIPS2017_6449f44a}.
We build the LtR training set by considering the last $50K$ queries in the training dataset used for building the graph. For each query, we consider all the candidate experts selected by \model. We remove from the set of queries those for which \model is not able to include, among the candidates, the actual expert who provided the accepted answer, and we extract, for each query and candidate expert pair, the features detailed in Section \ref{sec:ranking}. The resulting LtR training dataset includes 43,001 queries with 182 candidate experts each, on average. 
We split the train set into train and validation sets using the 0.8/0.2 splitting criteria. 
We performed hyper-parameter tuning with MRR on the validation set by exploiting the HyperOpt library~\cite{10.5555/3042817.3042832}. We optimized four learning parameters: \texttt{learning\_rate} $\in [0.0001, 0.15]$, \texttt{num\_leaves} $\in [50, 200]$, \texttt{n\_estimators} $\in [50, 150]$, \texttt{max\_depth} $\in [8, 15]$, and \texttt{min\_data\_in\_leaf} $\in [150, 500]$. 

\subsection{Baselines and competitors}
We compare \model with state-of-the-art competitors and variants of the proposed solution exploiting each only a subset of components. By examining these different configurations, we can assess and quantify the impact of each component and gain insights into the effectiveness of combining social and content information in our approach for addressing the  EF task for CQA platforms: 
\begin{itemize}
    \item \textbf{{BC}}: It uses the MLG and simply sorts the experts in the layers related to the new question based on their Betweenness centrality score.
    \item \textbf{{BM25}}: It uses the MLG and sorts the experts in the question's layers based on the BM25 score computed between the query and their previously answered questions. As in \model, the lists of the indexes are merged in alternating the elements.
    \item \textbf{{TUEF$_{\textbf{NB}}$}}: It uses the MLG and applies the Network-based method only. Candidates are ranked using a LtR model exploiting the static and the following query-dependent features: LayerCount, QueryKnowledge, VisitCountNetwork, StepsNetwork, BetweennesPos, BetweennessScore, Eigenvector, PageRank, Closeness, Degree, and AvgWeights.
    \item \textbf{{TUEF$_{\textbf{CB}}$}}: It uses the MLG and applies the Content-based method only. Candidates are ranked using a LtR model exploiting the static and the following query-dependent features: LayerCount, QueryKnowledge, VisitCountContent, StepsContent, ScoreIndexTag, ScoreIndexText, FrequencyIndexTag, FrequencyIndexTex, Degree, and AvgWeights.
    \item \textbf{{TUEF$_{\textbf{SL}}$}}: In contrast to \model, it represents users' relationships in a graph with a Single Layer. All other phases are unchanged.
    \item \textbf{{TUEF$_{\textbf{NoRW}}$}}: In contrast to \model, it skips the Exploratory phase and does not perform Random Walks to extend the set of candidate experts. 
    \item \textbf{\cite{roy2018finding}}: It ranks the experts according the solution of \cite{roy2018finding} which uses a linear combination of features modeling experts, questions, and users' expertise.
    \item \textbf{{NeRank}} \cite{li2019personalized}: It models the CQA platform as a heterogeneous network to learn representation for question raisers and question answerers through a metapath-based algorithm. With the heterogeneous network, NeRank preserves the information about relationships while it models the question's content with a single-layer LSTM. Finally, a CNN associates a score to each expert for a given question, representing the probability of the expert providing the accepted answer. 
\end{itemize}

\paragraph{\textbf{Evaluation metrics.}}
We test the model using the first 5,000 queries of the test set, which comprise only questions for which the best answerer is an expert.
We use Precision@1 (P@1), Normalized Discounted Cumulative Gain @3 (NDCG@3), Mean Reciprocal Rank (MRR), and Recall@100 (R@100) as our evaluation metrics and we compute statistical significance tests with the RanX Library \cite{DBLP:conf/ecir/Bassani22}.  
The cutoffs considered are low as for the EF task it is essential to find the relevant results at the top of the ranked lists. We mark statistically-significant performance gain/loss with respect to the best baseline (TUEF$_{CB}$)
with the symbols 
$\blacktriangle$ and 
$\blacktriangledown$ (paired $t$-test with $p$-value $< 0.05$). For  NeRank we report P@1, Hit@5 and MRR as these are the only metrics returned by the code. The source code of TUEF is publicly available\footnote{\url{https://github.com/maddalena-amendola/TUEF}}.

\vspace{-12pt}
\begin{table}
\caption{Comparison between \model and the baselines.
}
\centering
\rowcolors{2}{gray!15}{white}
\begin{tabular}{rcccc}
 & \multicolumn{1}{l}{\textbf{P@1}} & \multicolumn{1}{l}{\textbf{NDCG@3}} & \multicolumn{1}{l}{\textbf{R@100}} & \multicolumn{1}{l}{\textbf{MRR}} \\ 
 \toprule
{BC} & 0.020$^\blacktriangledown$ & 0.033$^\blacktriangledown$ & 0.076$^\blacktriangledown$ & 0.033$^\blacktriangledown$\\ 
{BM25} & 0.234$^\blacktriangledown$ &  0.355$^\blacktriangledown$ & 0.808$^\blacktriangledown$ & 0.369$^\blacktriangledown$\\ 
{TUEF$_{NB}$} & 0.066$^\blacktriangledown$ & 0.087$^\blacktriangledown$ & 0.213$^\blacktriangledown$ & 0.0930$^\blacktriangledown$ \\ 
\cite{roy2018finding} & 0.264$^\blacktriangledown$ & 0.359$^\blacktriangledown$ & 0.874$^\blacktriangle$ & 0.383$^\blacktriangledown$ \\
{TUEF$_{SL}$} & 0.436$^\blacktriangledown$ & 0.560$^\blacktriangledown$ & 0.826$^\blacktriangledown$ & 0.559$^\blacktriangledown$ \\ 
{TUEF$_{CB}$} & 0.447 & 0.573 & 0.849 &  0.572\\ 
{TUEF$_{NoRW}$} & 0.443 & 0.561$^\blacktriangledown$ & 0.754$^\blacktriangledown$ & 0.552$^\blacktriangledown$ \\
{\textbf{{\model}}} & \textbf{0.453} &\textbf{0.578}$^\blacktriangle$ & \textbf{0.874}$^\blacktriangle$ & \textbf{0.579}$^\blacktriangle$\\ 
\bottomrule
\end{tabular} 
\label{tab:results}
\vspace{-24pt}
\end{table}

\section{Results and Discussion}
\label{sec:results}

We first compare TUEF with all baselines but NeRank, as the latter requires some additional operations to make the approaches comparable, as explained later on.
Table \ref{tab:results} shows that \model consistently outperforms all baseline methods across all metrics considered. The best-performing competitor is \textit{$TUEF_{CB}$}, which relies only on the Content-based method. The improvements of \model over \textit{$TUEF_{CB}$} are statistically significant for all the metrics reported but P@1 where the two methods have similar performances even if \model has a slightly higher P@1. These findings emphasize the substantial contribution of the Content-based method to the model's performance, with the combined utilization of the Network-based method yielding a slightly superior performance. 
When comparing TUEF to the \textit{$TUEF_{NoRW}$} baseline, which stands as the best-performing baseline in terms of P@1, \model exhibits a relative improvement of 15.91\% for R@100. This result highlights the importance of the exploration of the MLG through Random Walks that identify candidate experts not retrieved with the other previous techniques. The P@1 value, however, suggests that the higher recall is not fully exploited by the \model ranking model, who seems not perfectly able to push the right candidate to the top position. 
Notably, the \textit{$TUEF_{NB}$} baseline, which solely employs the Network-based method, reports an R@100 of 0.213, merely a quarter of the R@100 achieved by the Content-based baseline.
The third most effective baseline, \textit{$TUEF_{SL}$}, is outperformed by \model by 3.9\% for P@1 and 5.81\% for R@100. This outcome highlights the advantages introduced by the \model MLG: the representation of experts under different macro topics enhance significantly the selection and the ranking of candidate experts.
Furthermore, even \textit{$TUEF_{CB}$}, the Content-based baseline, outperforms \textit{$TUEF_{SL}$}, emphasizing once more the significance of capturing question similarities and user relationships across distinct macro topics.
Instead, the method \cite{roy2018finding} shows comparable performance in terms of R@100, meaning that the linear combination of features successfully push the proper experts into the top-100 positions. However, the LtR techniques applied in \model allow to reach performance almost twice higher in terms of P@1. Using LtR techniques allows, in fact, to capture non-linear relations between the features, remarkably improving the ranking performance.
Lastly, the results of the \textit{$TUEF_{BC}$} and \textit{$TUEF_{BM25}$} baselines highlight that centrality measure alone is not a strong feature to choose the right expert, while who answered in the past the most similar question is the right expert in approximately 23\% of cases.
\vspace{-12pt}
\begin{figure}
    \centering        \includegraphics[width=0.6\textwidth]{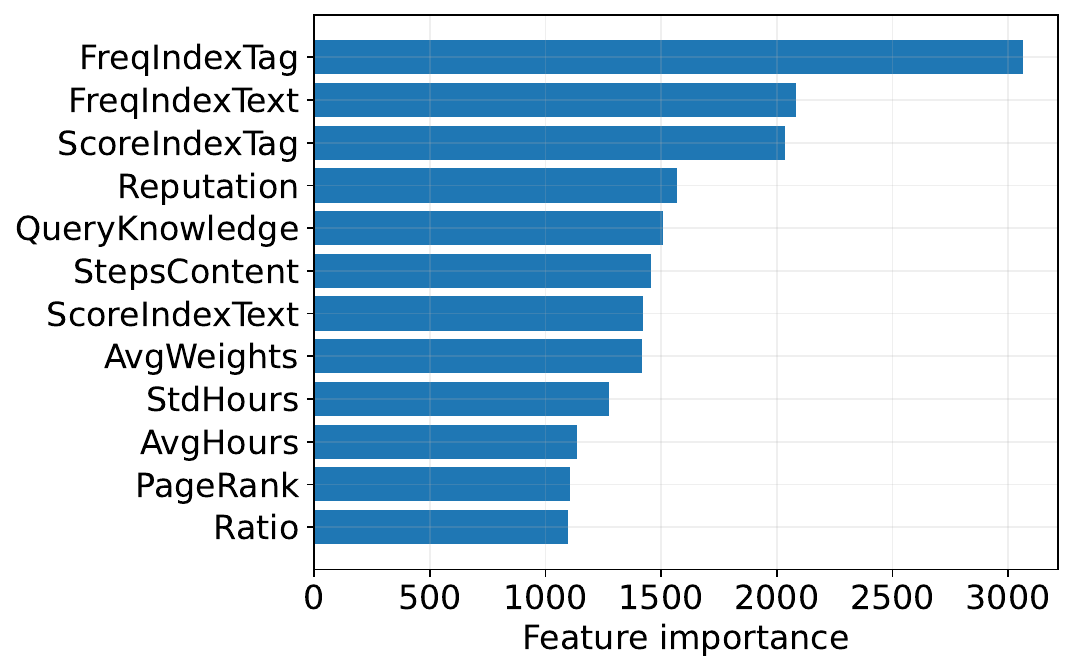}
    \caption{The feature importance as determined by the Learning-to-Rank approach.}
    \label{fig:feature_importance}
    \vspace{-24pt}
\end{figure}

\paragraph{\textbf{Feature Importance Analysis}}
Figure \ref{fig:feature_importance} highlights the twelve most influential factors in the \model algorithm's decision-making process, as determined by the LtR algorithm. Notably, \textit{FreqIndexTag} and \textit{FreqIndexText} stand out as the most significant features, emphasizing the importance of the frequency with which experts have addressed questions sourced from specific indexes.  Concurrently, \textit{ScoreIndexTag}, \textit{Reputation}, \textit{QueryKnowledge} and \textit{ScoreIndexText} indicate the importance given to the quality, relevance, and reliability of expert responses.
The presence of \textit{StepsContent}, \textit{AvgWeight}, and \textit{PageRank} reveal the role of network-related features, emphasizing both the path-based exploration and the structural importance of experts within the network. Additionally, the inclusion of \textit{StdHours} and \textit{AvgHours} features accentuates the algorithm's recognition of temporal activity patterns.

\vspace{-12pt}
\begin{table}[]
\caption{Comparison between \model and NeRank \cite{li2019personalized}. 
}
\label{tab:nerank}
\centering

\begin{tabular}{@{}lccc|ccc|ccc@{}}
 & \multicolumn{3}{c|}{StackOverflow} & \multicolumn{3}{c|}{Unix} & \multicolumn{3}{c}{AskUbuntu} \\ 
  \toprule
 & P@1 & Hit@5 & MRR & P@1 & Hit@5 & MRR & \multicolumn{1}{c}{P@1} & \multicolumn{1}{c}{Hit@5} & \multicolumn{1}{c}{MRR} \\ \cmidrule(l){2-10} 
NeRank & 0.34 & 0.708 & 0.498 & 0.568 & 0.882 & 0.698 & 0.499 & 0.89 & 0.676 \\
\textbf{TUEF} & \textbf{0.721}& \textbf{0.951} & \textbf{0.822} & \textbf{0.648}& \textbf{0.933} & \textbf{0.769} & \textbf{0.61} & \textbf{0.91} & \textbf{0.74}\\ \bottomrule
\end{tabular}
\label{tab:nerank}
\vspace{-12pt}
\end{table}

\paragraph{\textbf{Comparison with NeRank}}
NeRank adopts a different experimental setting. Specifically, for each test query $q$, NeRank generates and ranks a set of 20 candidate experts that include the $n$ experts that provided the answers to $q$, plus $20-n$ other experts randomly selected from the top 10\% most responsive users of the community. 
The apriori knowledge of the users answering the question from one side provides NeRank with a considerable advantage over the setting used for \model, but from the other side, it makes NeRank not suited for routing the query to the proper experts in the community. To fairly compare \model and NeRank, we applied the same experimental setting for \model. 
Specifically, for each query $q$, we consider, as in NeRank, the $n$ experts that answered $q$ plus $20-n$ other experts randomly selected from the ones collected during the \model exploratory phase. 
To highlight the contribution of \model, we experiment with two more StackExchange communities: Unix and AskUbuntu.
Given the high memory footprint and computational cost of NeRank, for all the tests we used only a sample of about 30,000 questions. 
Specifically, for StackOverflow, we consider the last 30,000 questions between those used in the previous experiments. Instead, for Unix and AskUbuntu, we consider the last 30,000 questions after extracting the data from January 2015 up to today. Given the small sample, a user is labeled as an expert if she has at least $\beta=5$ accepted answers. Table \ref{tab:nerank} reports the results comparing the two models considering the metrics returned by NeRank.
We can see that \model remarkably outperforms NeRank, with a relative improvement on StackOverflow of 65.06\% for MRR and exceeding 112\% for P@1. Moreover, the expert can be found in the top-5 ranked experts (Hit@5) in about 95.1\% of the \model predictions.
We can also notice an important improvement for the AskUbuntu community and an inferior one for the Unix community. This experimental methodology that uses the ground truth of the queried questions to identify the candidate experts remarkably facilitates the EF task as highlighted by the superior performance reported in Table \ref{tab:nerank} w.r.t. that in  Table \ref{tab:results}.

\section{Conclusions}
\label{sec:conclusions}

This work addressed the EF task in the context of  CQA platforms.
We contributed \model, a novel EF solution that combines content and social information available in CQA platforms. 
The empirical evaluation, performed on datasets from StackOverflow, demonstrates the superior performance of \model compared to existing methods. 
The improvements reach as high as 65.06\% for MRR and 112\% for P@1. By providing effective expert predictions, \model contributes to making CQA  engaging, trusted and credible.
In future work, we will investigate the interpretability and fairness of \model predictions.

\small{\paragraph{\textbf{Acknowledgements.}} This work was partially supported by: the H2020 SoBigData++ project (\#871042); the CAMEO PRIN project (\#2022ZLL7MW) funded by the MUR; the HEU EFRA project (\#101093026) funded by the EC under the NextGeneration EU programme. A. Passarella's and R. Perego's work was partly funded under the PNRR - M4C2 - Investimento 1.3,  PE00000013 - ``FAIR” project. However, the views and opinions expressed are those of the authors only and do not necessarily reflect those of the EU or European Commission-EU. Neither the EU nor the granting authority can be held responsible for them.}

\bibliographystyle{splncs04}
\bibliography{bib}

\end{document}